\documentclass{PoS}
\newcommand{\gsim}{\raisebox{-0.07cm   }
{$\, \stackrel{>}{{\scriptstyle\sim}}\, $}}

\usepackage{amsmath}
\usepackage{amssymb}
\usepackage{graphicx}
\usepackage{epstopdf}
\usepackage{float}
\usepackage{cite}
\usepackage{float}

\newcommand{\ep}{\varepsilon}
\newcommand{\Li}{\rm Li}
\newcommand{\N}{\nonumber}

\newcommand{\period}{\,.}
\newcommand{\comma}{\,,}
\newcommand{\GeV}{${\rm GeV}$}

\newcommand{\HSums}{{\tt HarmonicSums}}

\newcommand{\citeHSums}{\cite{Ablinger:2013hcp,Ablinger:2011te,Ablinger:2013cf}}

\newcommand{\CSsigma}{{\tt Sigma}}
\newcommand{\EvalMS}{{\tt EvaluateMultiSums}}
\newcommand{\SumProd}{{\tt SumProduction}}
\newcommand{\citeCSsigma}{ \cite{Schneider08JSC, Schneider05AC,
Schneider06JDEA, Schneider10AC, Schneider10PW, Schneider06SL,
Schneider07Hab, Ablinger:2010pb, Blumlein:2012hg}}
\newcommand{\citeEvalMS}{\cite{Blumlein:2012hg, Ablinger:2010ha, Schneider2013}}
\newcommand{\citeSumProd}{\cite{Blumlein:2012hg}}

\allowdisplaybreaks
\title{{\footnotesize DESY 13-223,~~~DO-TH 13/31,~~~MITP/13-073,~~~SFB/CPP-13-107,
~~~LPN 13-097}\\
New Results on the 3-Loop Heavy Flavor Corrections in Deep-Inelastic Scattering\thanks{
This work has been supported in part by DFG Sonderforschungsbereich
Transregio 9, Computergest\"utzte Theoretische Teilchenphysik, by Studienstiftung des Deutschen
Volkes, by the Austrian Science Fund (FWF) grants P20347-N18, SFB F50
(F5009-N15), by the EU Network {\sf LHCPHENOnet} PITN-GA-2010-264564, by the Reserach Center 
`Elementary Forces and Mathematical Foundations (EMG) of J. Gutenberg University Mainz and DFG, 
and by FP7 ERC Starting Grant  257638 PAGAP.}
}

\ShortTitle{New Results on the 3-Loop Heavy Flavor Corrections}

\author{A.~Behring, J.~Bl\"umlein, \speaker{A.~De Freitas}, T.~Pfoh, C.~Raab, M.~Round  
\\
        Deutsches Elektronen-Synchrotron DESY, Platanenallee 6, D-15738 Zeuthen, Germany\\
        E-mail: \email{abilio.de.freitas@desy.de}}

\author{J.~Ablinger, \speaker{A.~Hasselhuhn}, C.~Schneider, F.~Wi\ss{}brock\\
Research Institute for Symbolic Computation (RISC),\\
                          Johannes Kepler University, Altenbergerstra\ss{}e 69,
                          A--4040, Linz, Austria\\
E-mail: \email{alexander.hasselhuhn@risc.jku.at}}

\author{A.~von Manteuffel\\
        PRISMA Cluster of Excellence and Institute of Physics, J. Gutenberg University, 
D-55099 Mainz. Germany.}
\abstract{We report on recent progress in the calculation of the 3-loop massive Wilson 
coefficients in deep-inelastic scattering at general values of $N$ for neutral and charged 
current reactions in the asymptotic region $Q^2 \gg m^2$. Four new out of eight massive operator 
matrix elements and Wilson coefficients have been obtained recently. We also discuss recent results 
on Feynman graphs containing two-massive fermion lines and present complete results for the bubble
topologies for all processes.}

\FullConference{11th International Symposium on Radiative Corrections
(Applications of Quantum Field Theory to Phenomenology) (RADCOR 2013),\\
                22-27 September 2013\\
                Lumley Castle Hotel, Durham, UK }

\begin{document}
%-------------------------------------------------------------------------------------------------
%-------------------------------------------------------------------------------------------------
\section{Introduction}
%-------------------------------------------------------------------------------------------------

\noindent
The Wilson coefficients for the heavy quark contributions are known to 2-loop order 
in semi-analytic form
\cite{Laenen:1992zk,Bierenbaum:2009zt}.\footnote{For a precise numerical implementation in
Mellin space see \cite{Alekhin:2003ev}.} In the asymptotic region of large virtualities $Q^2 \gg m^2$, 
the Wilson coefficients were calculated analytically in 
\cite{Buza:1995ie,Buza:1996wv,Bierenbaum:2007dm,Bierenbaum:2007qe,Bierenbaum:2008yu}. This 
approximation 
holds in case of $F_2(x,Q^2)$ for scales of $Q^2/m^2 \gsim 10$ at the 1\% level \cite{Buza:1995ie}. 
In 2009 a series of Mellin moments $N = 2 ... 10 (12,14)$ was calculated for all massive operator 
matrix elements (OMEs) in \cite{Bierenbaum:2009mv} mapping these moments to massive tadpoles, which 
could be calculated using {\tt MATAD}, \cite{Steinhauser:2000ry}. In the asymptotic region also the 
3-loop corrections 
for $F_L(x,Q^2)$ were computed \cite{Blumlein:2006mh}, which are, however, only applicable at much higher
scales. 

Including the case of transversity \cite{Blumlein:2009rg} there are eight unpolarized massive Wilson 
coefficients at three loop order to be calculated. In 2010 the Wilson coefficients $L_{qg,Q}^{(3)}$ and 
$L_{qq,Q}^{(3),\rm PS}$ were computed \cite{Ablinger:2010ty}. Here the exchanged gauge boson couples
to a massless fermion line. Furthermore, all contributions due to the color factors $C_{F,A} T_F^2 
N_F$ have been computed in \cite{Ablinger:2010ty,Blumlein:2012vq}. 3-loop ladder and $V$-topologies 
have been studied in detail in \cite{Ablinger:2012qm,Ablinger:2012sm}. In all these calculations 
after 
performing the Feynman parameter integrals, nested finite and infinite sums over hypergeometric 
terms,
cf.~\cite{Blumlein:2010zv}, occur, which have to be solved by applying modern summation 
technologies.\footnote{For a recent survey article of the summation packages see
see Ref.~\cite{Schneider:2013zna}.} These are 
encoded in the packages \CSsigma \citeCSsigma, \EvalMS \citeEvalMS, \SumProd \citeSumProd, and 
$\rho${\tt-Sum}~\cite{ROUND}. Algebraic and structural relations between sums of specific types, such 
as 
harmonic sums \cite{HSUM,Blumlein:2003gb,Blumlein:2009ta}, multiple zeta values \cite{Blumlein:2009cf}, 
harmonic polylogarithms \cite{Remiddi:1999ew}, generalized harmonic sums and associated polylogarithms 
\cite{Moch:2001zr,Ablinger:2013cf}, cyclotomic harmonic sums and polylogarithms 
\cite{Ablinger:2011te}, 
as well as binomially weighted finite sums and the associated polylogarithms \cite{RAAB}, are 
mutually applied in these calculations.\footnote{For a recent review see \cite{Ablinger:2013jta}.} 
The corresponding relations and algorithms are encoded in the package {\tt HarmonicSums} 
\cite{Ablinger:2013cf,Ablinger:2013hcp}.

During the last year we have calculated four more OMEs at three loop order, 
$A_{qq,Q}^{(3), \rm NS},~A_{qq,Q}^{(3), \rm NS,TR},~A_{gq,Q}^{(3)}$ and $A_{Qq}^{(3), \rm PS}$ 
and the associated massive Wilson coefficients for  $A_{qq,Q}^{(3), \rm NS}$ and $A_{Qq}^{(3), \rm PS}$ 
at large $Q^2$. The corresponding topologies were first reduced to master integrals using 
integration-by-parts relations \cite{IBP} using the package {\tt Reduze2} 
\cite{Studerus:2009ye,vonManteuffel:2012np}. The 
master integrals were finally calculated using different summation technologies being described 
above. Furthermore, we computed the bubble topologies for all OMEs containing one massless bubble. 
Progress has also been made in the calculation of the topologies with two massive lines of the same 
mass.

In this note we report on these series of results obtained recently. In Section 2 the yet missing 
results for the 2-bubble topologies, with one massless line beyond the results given in 
\cite{Ablinger:2010ty,Ablinger:2012qm}, are presented. In Section 3 we discuss complete results 
obtained for four new massive OMEs and Wilson coefficients. Results on graphs with two massive quark 
lines of equal masses are discussed in Section 4. The asymptotic massive two-loop Wilson coefficients 
for charged current reactions are given in Section 5, and Section 6 contains the conclusions. 
%%%%%%%%%%%%%%%%%%%%%%%%%%%%%%%%%%%%%%%%%%%%%%%%%%%%%%%%%%%%%%%%%%%%%%%%%%%%%%%%%%%%%%%%%%%%%%%%%%
\section{Results for Bubble Graphs}
%%%%%%%%%%%%%%%%%%%%%%%%%%%%%%%%%%%%%%%%%%%%%%%%%%%%%%%%%%%%%%%%%%%%%%%%%%%%%%%%%%%%%%%%%%%%%%%%%%

\noindent
All 2-bubble topologies have been calculated for all the massive operator matrix elements
at general $N$. The contributions $\propto N_F T_F^2$ for $A_{Qq}^{(3) \rm PS}$ and $A_{Qg}^{(3)}$
have been obtained before in \cite{Ablinger:2010ty} and for $A_{gg}^{(3)}$ in \cite{Blumlein:2012vq}.
Likewise, the terms $\propto T_F^2$ were given in \cite{Ablinger:2011pb} for $A_{Qq}^{(3) \rm 
PS}$.\footnote{In (\ref{eq:BUB1}) We corrected a typograpgical error in \cite{Ablinger:2011pb}.}  
In the following we list the 
corresponding results for $a_{ij}^{k, {\sf b}}$ in the pure-singlet-, $gg$- and $Qg$-cases. Here 
harmonic sums $S_{\vec{a}}(N) \equiv S_{\vec{a}}$ 
up to weight {\sf w = 5}, including negative indices contribute. The calculation of the 
corresponding 
graphs has been carried out directly using (generalized) hypergeometric function techniques for the 
whole diagrams 
to convert them into sum-representations. The latter were solved using modern 
summation techniques
 as 
encoded in the packages \CSsigma \citeCSsigma, \EvalMS \citeEvalMS, \SumProd \citeSumProd, and
$\rho${\tt-Sum}~\cite{ROUND}. 

For the pure-singlet OME all contributions are given. The constant part of the unrenormalized OME
reads~:
%----------------------------------------------------------------------------------------
\begin{eqnarray}
\label{eq:BUB1}
a_{Qq}^{\rm PS,{\sf b}}(N) &=&
%--------
\textcolor{blue}{C_F^2 T_F} \frac{1}{(N-1) N^2}
\Biggl\{
%----
\frac{4 P_{22}}{3 (N-1)^3 N^3 (N+1)^5 (N+2)^4}
%----
-\frac{24 \big(N^2+N+2\big)^2}{(N+1)^2 (N+2)} S_3
   \nonumber \\ &&
%----
+\frac{2 P_9}{(N-1) N (N+1)^3 (N+2)^2} \big[\zeta_2 + 4 S_2\big]
%----
+\frac{28 \big(N^2+N+2\big)^2 \zeta_3}{3 (N+1)^2 (N+2)}
%----
\Biggr\}
   \nonumber \\ &&
%--------
+\textcolor{blue}{C_F T_F^2}
\Biggl\{
%----
\frac{1}{(N-1) N^2}
\Biggl[
\frac{32 \big(N^2+N+2\big)^2 }{27 (N+1)^2 (N+2)}S_1^2
-\frac{160 \big(N^2+N+2\big)^2 }{9 (N+1)^2 (N+2)}S_2
   \nonumber \\ &&
+\frac{64 P_{18}}{81 N^2 (N+1)^4 (N+2)^3 (N+3) (N+4) (N+5)}
\Biggr] S_1
   \nonumber \\ &&
%----
-\frac{64 P_{21}}{243 (N-1) N^5 (N+1)^4 (N+2)^4 (N+3) (N+4) (N+5)}
   \nonumber \\ &&
%----
+\frac{32}{27 (N-1) N^3 (N+1)^2 (N+2)^2 (N+3) (N+4) (N+5)} \left[P_{12} S_2 - \frac{P_{14} 
S_1^2}{N+1}\right]
   \nonumber \\ &&
%----
-\frac{512 \big(N^2+N+2\big)^2 }{27 (N-1) N^2 (N+1)^2 (N+2)}S_3
%----
+\frac{128 \big(N^2+N+2\big)^2}{3 (N-1) N^2 (N+1)^2 (N+2)} S_{2,1}
   \nonumber \\ &&
%----
+\Biggl[
\frac{32 \big(N^2+N+2\big)^2 }{3 (N-1) N^2 (N+1)^2 (N+2)}S_1
-\frac{32 P_2}{9 (N-1) N^3 (N+1)^2 (N+2)^2}
\Biggr] \zeta_2
   \nonumber \\ &&
%----
-\frac{1024 \big(N^2+N+2\big)^2 \zeta_3}{9 (N-1) N^2 (N+1)^2 (N+2)}
%----
\Biggr\}
%--------
+\textcolor{blue}{C_F T_F^2 n_f}
\Biggl\{
%----
-\frac{16 \big(N^2+N+2\big)^2 }{27 (N-1) N^2 (N+1)^2 (N+2)}S_1^3
   \nonumber \\ &&
%----
+\frac{16 P_5 }{27 (N-1) N^3 (N+1)^3 (N+2)^2}S_1^2
%----
+\Biggl[
-\frac{208 \big(N^2+N+2\big)^2}{9 (N-1) N^2 (N+1)^2 (N+2)} S_2
   \nonumber \\ &&
-\frac{32 P_{15}}{81 (N-1) N^4 (N+1)^4 (N+2)^3}
\Biggr] S_1
%----
+\frac{32 P_{19}}{243 (N-1) N^5 (N+1)^5 (N+2)^4}
   \nonumber \\ &&
%----
+\frac{208 P_5 }{27 (N-1) N^3 (N+1)^3 (N+2)^2}S_2
%----
-\frac{1760 \big(N^2+N+2\big)^2 }{27 (N-1) N^2 (N+1)^2 (N+2)}S_3
   \nonumber \\ &&
%----
+\Biggl[
\frac{16 P_5}{9 (N-1) N^3 (N+1)^3 (N+2)^2}
-\frac{16 \big(N^2+N+2\big)^2 }{3 (N-1) N^2 (N+1)^2 (N+2)}S_1
\Biggr] \zeta_2
   \nonumber \\ &&
%----
+\frac{224 \big(N^2+N+2\big)^2 \zeta_3}{9 (N-1) N^2 (N+1)^2 (N+2)}
%----
\Biggr\}
%--------
+\textcolor{blue}{C_A C_F T_F}
\Biggl\{
%----
\frac{\big(2 N^3+5 N^2-14 N-24\big) }{72 N^2 (N+1)^2 (N+2)}S_1^4
   \nonumber \\ &&
%----
+\frac{P_6 }{54 (N-1) N^3 (N+1)^3 (N+2)^2}S_1^3
%----
+\Biggl[
\frac{P_{13}}{54 (N-1) N^4 (N+1)^4 (N+2)^3}
   \nonumber \\ &&
+\frac{\big(74 N^4+523 N^3+733 N^2+374 N+440\big) }{12 (N-1) N^2 (N+1)^2 (N+2)}S_2
\Biggr] S_1^2
%----
+\Biggl[
\frac{P_{20}}{81 (N-1) N^5 (N+1)^5 (N+2)^4}
   \nonumber \\ &&
+\frac{P_7 }{18 (N-1) N^3 (N+1)^3 (N+2)^2}S_2
+\frac{\big(134 N^4+1971 N^3+3857 N^2+2270 N+936\big) }{9 (N-1) N^2 (N+1)^2 (N+2)}S_3
   \nonumber \\ &&
-\frac{2 \big(4 N^4+107 N^3+255 N^2+122 N-32\big) }{3 (N-1) N^2 (N+1)^2 (N+2)}S_{2,1}
-\frac{64 \big(N^3+6 N^2+3 N+2\big) }{3 (N-1) N^2 (N+1)^2}S_{-2,1}
\Biggr] S_1
   \nonumber \\ &&
%----
+\frac{\big(82 N^4+1211 N^3+3061 N^2+2982 N+2008\big) }{24 (N-1) N^2 (N+1)^2 (N+2)}S_2^2
   \nonumber \\ &&
%----
+\frac{16 \big(N^4+2 N^3+7 N^2+6 N+4\big) }{3 (N-1) N^2 (N+1)^2 (N+2)}S_{-2}^2
%----
+\frac{P_{23}}{243 (N-1)^5 N^5 (N+1)^6 (N+2)^5}
   \nonumber \\ &&
%----
+\Biggl[
\frac{8 P_4}{3 (N-1) N^3 (N+1)^3 (N+2)^2}
+\frac{32 \big(N^3+6 N^2+3 N+2\big) }{(N-1) N^2 (N+1)^2}S_1
\Biggr] S_{-3}
   \nonumber \\ &&
%----
+\frac{P_{17}}{54 (N-1)^3 N^4 (N+1)^4 (N+2)^3} S_2
%----
+\Biggl[
\frac{32 \big(N^3+6 N^2+3 N+2\big) }{3 (N-1) N^2 (N+1)^2}S_1^2
   \nonumber \\ &&
+\frac{32 P_1 }{3 (N-1) N^3 (N+1)^3 (N+2)}S_1
-\frac{4 P_{11}}{3 (N-1) N^4 (N+1)^4 (N+2)^3}
   \nonumber \\ &&
+\frac{16 \big(5 N^4+58 N^3+99 N^2+46 N+20\big) }{3 (N-1) N^2 (N+1)^2 (N+2)}S_2
\Biggr] S_{-2}
%----
+\frac{P_{10}}{27 (N-1)^2 N^3 (N+1)^3 (N+2)^2} S_3
   \nonumber \\ &&
%----
+\frac{\big(194 N^4+4719 N^3+10489 N^2+6814 N+2136\big) }{12 (N-1) N^2 (N+1)^2 (N+2)}S_4
   \nonumber \\ &&
%----
+\frac{32 \big(3 N^4+36 N^3+61 N^2+28 N+12\big) }{3 (N-1) N^2 (N+1)^2 (N+2)}S_{-4}
%----
-\frac{2 P_3 }{3 (N-1) N^3 (N+1)^3 (N+2)^2}S_{2,1}
   \nonumber \\ &&
%----
-\frac{2 \big(293 N^3+813 N^2+470 N-80\big) }{3 (N-1) N^2 (N+1)^2 (N+2)}S_{3,1}
%----
-\frac{32 P_1 }{3 (N-1) N^3 (N+1)^3 (N+2)}S_{-2,1}
   \nonumber \\ &&
%----
-\frac{32\big(N^3+6 N^2+3 N+2\big) }{3 (N-1) N^2 (N+1)^2} \big[2 S_{-2,2} + 3 S_{-3,1}\big]
%----
+\frac{64 \big(N^3+6 N^2+3 N+2\big) }{3 (N-1) N^2 (N+1)^2}S_{-2,1,1}
   \nonumber \\ &&
%----
-\frac{2 \big(2 N^4-97 N^3-267 N^2-118 N+88\big) }{3 (N-1) N^2 (N+1)^2 (N+2)}S_{2,1,1}
%----
+\Biggl[
\frac{\big(2 N^3+5 N^2-14 N-24\big) }{4 N^2 (N+1)^2 (N+2)}S_1^2
   \nonumber \\ &&
+\frac{P_6 }{6 (N-1) N^3 (N+1)^3 (N+2)^2}S_1
+\frac{P_{16}}{18 (N-1)^3 N^3 (N+1)^4 (N+2)^3}
   \nonumber \\ &&
+\frac{\big(2 N^4+95 N^3+265 N^2+222 N+88\big) }{4 (N-1) N^2 (N+1)^2 (N+2)}S_2
+\frac{4 \big(N^4+14 N^3+23 N^2+10 N+4\big) }{(N-1) N^2 (N+1)^2 (N+2)}S_{-2}
\Biggr] \zeta_2
   \nonumber \\ &&
%----
+\Biggl[
\frac{P_8}{9 (N-1)^2 N^3 (N+1)^3 (N+2)^2}
+\frac{\big(34 N^3+109 N^2+98 N-24\big) }{3 N^2 (N+1)^2 (N+2)}S_1
\Biggr] \zeta_3
%----
\Biggr\}~.
%--------
\end{eqnarray}
%----------------------------------------------------------------------------------------
The polynomials $P_i$ read~:
%----------------------------------------------------------------------------------------
\begin{eqnarray}
P_1&=&N^6+11 N^5+64 N^4+87 N^3+33 N^2+16 N+4 \\
P_2&=&8 N^6+29 N^5+84 N^4+193 N^3+162 N^2+124 N+24 \\
P_3&=&2 N^7+65 N^6+591 N^5+1904 N^4+2554 N^3+1132 N^2-120 N-48 \\
P_4&=&4 N^7+67 N^6+505 N^5+1277 N^4+1227 N^3+476 N^2+156 N+16 \\
P_5&=&8 N^7+37 N^6+68 N^5-11 N^4-86 N^3-56 N^2-104 N-48 \\
P_6&=&43 N^7+221 N^6+694 N^5+722 N^4-496 N^3-424 N^2+464 N+96 \\
P_7&=&631 N^7+3965 N^6+17170 N^5+33194 N^4+22160 N^3+5912 N^2+5456 N+864 \\
P_8&=&-331 N^8-1540 N^7-3434 N^6-1648 N^5+4089 N^4+1452 N^3-2316 N^2 \nonumber \\
    &&+560 N+1152 \\
P_9&=&2 N^8+6 N^7-N^6-51 N^5-59 N^4-19 N^3-26 N^2+28 N+24 \\
P_{10}&=&2365 N^8+13228 N^7+55085 N^6+90910 N^5+4596 N^4-91944 N^3-55632 N^2 \nonumber \\
       &&-2960 N-96 \\
P_{11}&=&N^9-215 N^8-2293 N^7-7913 N^6-12020 N^5-8528 N^4-3048 N^3-848 N^2 \nonumber \\
       &&+96 N+128 \\
P_{12}&=&40 N^9+625 N^8+3284 N^7+5392 N^6-7014 N^5-33693 N^4-47454 N^3 \nonumber \\
       &&-46100 N^2-26280 N+7200 \\
P_{13}&=&-359 N^{10}-2734 N^9-9528 N^8-14379 N^7-11852 N^6-28608 N^5-46716 N^4 \nonumber \\
       &&-8528 N^3+22240 N^2+7296 N+576 \\
P_{14}&=&8 N^{10}+133 N^9+1095 N^8+5724 N^7+18410 N^6+34749 N^5+40683 N^4 \nonumber \\
       &&+37370 N^3+22748 N^2-3960 N-7200 \\
P_{15}&=&25 N^{10}+176 N^9+417 N^8+30 N^7-20 N^6+1848 N^5+2244 N^4+1648 N^3 \nonumber \\
       &&+3040 N^2+2112 N+576 \\
P_{16}&=&-359 N^{11}-2025 N^{10}-4518 N^9+2510 N^8+21229 N^7+14611 N^6-14384 N^5 \nonumber \\
       &&-16352 N^4-6592 N^3+152 N^2+6784 N+4128 \\
P_{17}&=&-4739 N^{12}-27252 N^{11}-62919 N^{10}+29003 N^9+277786 N^8+167821 N^7 \nonumber \\
       &&-215504 N^6-163112 N^5-31660 N^4-19696 N^3+67744 N^2+46464 N
-1728 \\
P_{18}&=&52 N^{13}+746 N^{12}+4658 N^{11}+20431 N^{10}+79990 N^9+251778 N^8+553796 N^7 \nonumber \\
       &&+837697 N^6+886552 N^5+599060 N^4+155864 N^3-82368 N^2-76896 N\nonumber\\ &&
-17280 \\
P_{19}&=&158 N^{13}+1663 N^{12}+7714 N^{11}+23003 N^{10}+56186 N^9+89880 N^8+59452 N^7 \nonumber \\
       &&-8896 N^6-12856 N^5-24944 N^4-84608 N^3-77952 N^2-35712 N-6912 \\
P_{20}&=&1474 N^{13}+15137 N^{12}+67586 N^{11}+156550 N^{10}+284233 N^9+530832 N^8 \nonumber \\
       &&+412460 N^7-695900 N^6-1291340 N^5-157480 N^4+639968 N^3+318720 N^2 \nonumber \\
       &&+72000 N+3456 \\
P_{21}&=&293 N^{15}+4670 N^{14}+32280 N^{13}+145948 N^{12}+559575 N^{11}+1871440 N^{10} \nonumber \\
       &&+4877344 N^9+9333994 N^8+12958212 N^7+12693884 N^6+8472792 N^5 \nonumber \\
       &&+4514336 N^4+3109248 N^3+2192832 N^2+1026432 N+207360 \\
P_{22}&=&4 N^{16}+30 N^{15}+101 N^{14}+301 N^{13}+561 N^{12}-1385 N^{11}-4474 N^{10}+324 N^9 \nonumber \\
       &&+4667 N^8-4115 N^7-2529 N^6+6629 N^5-330 N^4-3672 N^3-1024 N^2 \nonumber \\
       &&+976 N+480 \\
P_{23}&=&-8780 N^{19}-83054 N^{18}-302761 N^{17}-396603 N^{16}+104969 N^{15}+2043037 N^{14} \nonumber \\
       &&+5908471 N^{13}+1725207 N^{12}-16095317 N^{11}-11836443 N^{10}+21978990 N^9 \nonumber \\
       &&+16243568 N^8-18166796 N^7-7483912 N^6+11581992 N^5+1162152 N^4 \nonumber \\
       &&-5841152 N^3-833088 N^2+1415808 N+563328~.
\end{eqnarray}
%----------------------------------------------------------------------------------------

The new contributions to the finite part of the OME $A_{gg}^{(3)}$ read~:
%----------------------------------------------------------------------------------------
\begin{eqnarray}
a_{gg}^{\sf b}(N) &=&
%--------
\textcolor{blue}{T_F C_A^2}
\Biggl\{
-\frac{13}{36} S_1^5
%--
+\frac{R_{14}}{864 (N-1) N (N+1) (N+2)} S_1^4
%--
+\Biggl[
\frac{R_{20}}{1944 (N-1)^2 N^2 (N+1)^2 (N+2)^2}
   \nonumber \\ &&
-\frac{235 }{54}S_2
\Biggr] S_1^3
%--
+\Biggl[
\frac{R_{29}}{648 (N-1)^3 N^3 (N+1)^3 (N+2)^3}
+\frac{R_{15}}{432 (N-1) N (N+1) (N+2)} S_2
   \nonumber \\ &&
-5 S_3+\frac{4}{9} S_{2,1}
+\frac{32}{9} S_{-2,1}
\Biggr] S_1^2
%--
+\Biggl[
-\frac{49}{12} S_2^2
+\frac{R_{21}}{216 (N-1)^2 N^2 (N+1)^2 (N+2)^2} S_2
   \nonumber \\ &&
+\frac{R_{30}}{972 (N-1)^4 N^4 (N+1)^4 (N+2)^4}
+\frac{R_{13}}{108 (N-1) N (N+1) (N+2)} S_3
-\frac{179 }{18}S_4
   \nonumber \\ &&
+\frac{R_6 }{27 (N-1) N (N+1) (N+2)}S_{2,1}
-\frac{32}{9} S_{3,1}
+\frac{64 (5 N+22)}{27 (N+2)} S_{-2,1}
+\frac{32}{9} S_{-2,2}
-\frac{8}{9} S_{2,1,1}
   \nonumber \\ &&
-\frac{128}{9} S_{-2,1,1}
\Biggr] S_1
%--
+\frac{R_{16}}{864 (N-1) N (N+1) (N+2)} S_2^2
%--
   \nonumber \\ &&
+\frac{R_{32}}{46656 (N-1)^5 N^5 (N+1)^5 (N+2)^5}
%--
+\frac{R_{22}}{972 (N-1)^2 N^2 (N+1)^2 (N+2)^2} S_3
%--
   \nonumber \\ &&
+\frac{R_{12}}{432 (N-1) N (N+1) (N+2)} S_4
%--
+6 S_5
%--
+\frac{448}{81} S_{-3}
%--
-\frac{80}{27} S_{-4}
%--
+\frac{16}{9} S_{-5}
%--
   \nonumber \\ &&
+\frac{R_{17}}{81 (N-1) N (N+1)^2 (N+2)^2} S_{2,1}
%--
+ \Biggl[
-\frac{16}{27} S_1^3
-\frac{64 (N+3) (41 N+56)}{81 (N+1) (N+2)}
   \nonumber \\ &&
-\frac{16 (5 N+22) }{27 (N+2)}S_1^2
+\Biggl(\frac{16 }{9}S_2-\frac{128 \big(7 N^2+36 N+38\big)}{81 (N+1) (N+2)}\Biggr) S_1
+\frac{16 (5 N+22) }{27 (N+2)}S_2
-\frac{32 }{27}S_3
   \nonumber \\ &&
+\frac{64}{9} S_{2,1}
\Biggr] S_{-2}
%--
+\frac{20}{9} S_{2,3}
%--
+\frac{64}{9} S_{2,-3}
%--
+\frac{R_3 }{27 (N-1) N (N+1) (N+2)}S_{3,1}
%--
-\frac{34}{9} S_{4,1}
   \nonumber \\ &&
%--
+ \Biggl[
\frac{R_{28}}{648 (N-1)^3 N^3 (N+1)^3 (N+2)^3}
-\frac{349 }{27}S_3
-\frac{8}{9} S_{2,1}
-\frac{32}{3} S_{-2,1}
\Biggl] S_2
   \nonumber \\ &&
%--
+\frac{256 \big(7 N^2+36 N+38\big)}{81 (N+1) (N+2)} S_{-2,1}
%--
+\frac{32 (5 N+22)}{27 (N+2)} S_{-2,2}
%--
+\frac{R_1 }{27 (N-1) N (N+1) (N+2)}S_{2,1,1}
   \nonumber \\ &&
-\frac{64}{9} S_{-2,3}
%--
%--
-\frac{64}{9} S_{2,1,-2}
%--
-\frac{4}{9} S_{2,2,1}
%--
+\frac{52}{9} S_{3,1,1}
%--
-\frac{128 (5 N+22)}{27 (N+2)} S_{-2,1,1}
%--
-\frac{64}{9} S_{-2,2,1}
   \nonumber \\ &&
%--
+\frac{64}{9} S_{2,1,1,1}
%--
+\frac{256}{9} S_{-2,1,1,1}
%--
+\Biggl[
-\frac{13}{6} S_1^3
+\frac{R_{11} S_1^2}{48 (N-1) N (N+1) (N+2)}
   \nonumber \\ &&
+\Biggl(\frac{R_{18}}{72 (N-1) N^2 (N+1)^2 (N+2)^2}-\frac{55 }{6}S_2\Biggr) S_1
+\frac{R_{26}}{864 (N-1)^3 N^3 (N+1)^3 (N+2)^3}
   \nonumber \\ &&
+ \Biggl(-\frac{2 (5 N+18)}{3 (N+2)}  -\frac{4 }{3}S_1\Biggr) S_{-2}
+\frac{R_9 }{48 (N-1) N (N+1) (N+2)}S_2
+\frac{8 }{3}S_3+\frac{2}{3} S_{-3}
   \nonumber \\ &&
+\frac{5}{3} S_{2,1}
+\frac{8}{3} S_{-2,1}
\Biggr] \zeta_2
%--
+\Biggl[
-\frac{49}{3} S_1^2
+\frac{7 R_{10} S_1}{36 (N-1) N (N+1) (N+2)}
+7 S_2
+\frac{14}{3} S_{-2}
   \nonumber \\ &&
-\frac{7 R_{25}}{216 (N-1)^2 N^2 (N+1)^2 (N+2)^2}
\Biggr] \zeta_3
%--
\Biggr\}~.
\end{eqnarray}
%----------------------------------------------------------------------------------------
The polynomials $R_i$ are
%----------------------------------------------------------------------------------------
\begin{eqnarray}
R_1&=&-353 N^4-832 N^3+443 N^2+934 N+96 \\
%%R_2&=&4 N^4+19 N^3+48 N^2+47 N+62 \\
R_3&=&7 N^4-280 N^3-277 N^2-26 N-288 \\
%%R_4&=&20 N^4-35 N^3-326 N^2-361 N-378 \\
%%R_5&=&40 N^4+99 N^3+30 N^2-11 N+82 \\
R_6&=&121 N^4+368 N^3-211 N^2-470 N-96 \\
%%R_7&=&140 N^4+391 N^3+358 N^2+269 N+642 \\
%%R_8&=&320 N^4+643 N^3-398 N^2-775 N-150 \\
R_9&=&343 N^4+1036 N^3+2033 N^2+2908 N+3316 \\
R_{10}&=&351 N^4+824 N^3+385 N^2+456 N+1068 \\
R_{11}&=&815 N^4+1932 N^3+537 N^2+60 N+1684 \\
R_{12}&=&1957 N^4+14588 N^3+38867 N^2+50236 N+55020 \\
R_{13}&=&3259 N^4+9004 N^3+4069 N^2+980 N+8596 \\
R_{14}&=&3311 N^4+8148 N^3+4041 N^2+3204 N+9508 \\
R_{15}&=&6869 N^4+16876 N^3+18403 N^2+22892 N+33420 \\
R_{16}&=&12653 N^4+32092 N^3+4987 N^2-8404 N+20268 \\
R_{17}&=&1270 N^6+8222 N^5+16333 N^4+7130 N^3-7481 N^2-2050 N+2928 \\
R_{18}&=&-1571 N^7-9661 N^6-26791 N^5-49153 N^4-67528 N^3-55096 N^2 
-11384 N
\nonumber\\
       &&
+8064 \\
%%R_{19}&=&1080 N^7+3257 N^6-1075 N^5-7371 N^4+1579 N^3+6718 N^2-156 N-792 \\
R_{20}&=&-40553 N^8-185774 N^7-259150 N^6-122366 N^5-173461 N^4-129392 N^3 \nonumber\\
       &&+366064 N^2+293208 N-59616 \\
R_{21}&=&-17939 N^8-83762 N^7-104386 N^6+5070 N^5-24959 N^4-144464 N^3 \nonumber\\
       &&+25472 N^2+106408 N-3360 \\
R_{22}&=&-3185 N^8-62618 N^7-314842 N^6-626834 N^5-451705 N^4+424288 N^3 \nonumber\\
       &&+1063528 N^2+416328 N-133920 \\
%%R_{23}&=&13 N^8+165 N^7+901 N^6+1778 N^5+1387 N^4-1231 N^3-3921 N^2-1504 N+252 \\
%%R_{24}&=&360 N^8+1589 N^7+1850 N^6-238 N^5+936 N^4+2917 N^3-2186 N^2-2804 N-264 \\
R_{25}&=&763 N^8+4224 N^7+10030 N^6+18476 N^5+22927 N^4+8348 N^3-9888 N^2 \nonumber\\
       &&-8944 N+5904 \\
R_{26}&=&2977 N^{12}+27302 N^{11}+121749 N^{10}+400754 N^9+654511 N^8-19518 N^7-1184809 N^6 \nonumber\\
       &&-1028282 N^5-42444 N^4-316832 N^3-1364944 N^2-481248 N+120384 \\
%%R_{27}&=&3472 N^{12}+21666 N^{11}+37911 N^{10}-15601 N^9-89343 N^8-10653 N^7+90581 N^6 \nonumber\\
%%       &&-4713 N^5-75303 N^4-3383 N^3+24330 N^2+2388 N-792 \\
R_{28}&=&23572 N^{12}+191008 N^{11}+720025 N^{10}+1584161 N^9+1802946 N^8-206032 N^7 \nonumber\\
       &&-3384669 N^6-3005177 N^5+1544742 N^4+2995680 N^3+265168 N^2 \nonumber\\
       &&-238944 N+195840 \\
R_{29}&=&96724 N^{12}+618064 N^{11}+1208881 N^{10}+170137 N^9-1519638 N^8-213008 N^7 \nonumber\\
       &&+897211 N^6-1994577 N^5-2283122 N^4+584016 N^3+268880 N^2-319584 N \nonumber\\
       &&-2304 \\
R_{30}&=&-290424 N^{16}-2538872 N^{15}-8370109 N^{14}-11564871 N^{13}-1339582 N^{12} \nonumber\\
       &&+15569491 N^{11}+25755378 N^{10}+32901569 N^9+17718754 N^8-29577639 N^7 \nonumber\\
       &&-33515273 N^6+12365378 N^5+17499480 N^4+73824 N^3-625344 N^2+266688 N \nonumber\\
       &&-736128 \\
%%R_{31}&=&18713 N^{16}+227024 N^{15}+1245276 N^{14}+3564368 N^{13}+4664666 N^{12}-1051728 N^{11} \nonumber\\
%%       &&-12936804 N^{10}-17722464 N^9-3014511 N^8+17976256 N^7+15522400 N^6 \nonumber \\
%%       &&-4626384 N^5-10818812 N^4-3994592 N^3-471456 N^2+70272 N+150336 \\
R_{32}&=&3752873 N^{20}+44889498 N^{19}+231324635 N^{18}+699986798 N^{17}+1323895202 N^{16} \nonumber\\
       &&+1047978356 N^{15}-2040696426 N^{14}-6981260596 N^{13}-6762635091 N^{12} \nonumber\\
       &&+2129282098 N^{11}+9092892879 N^{10}+5284986214 N^9-791167784 N^8-3464256800 N^7 \nonumber\\
       &&-6299611472 N^6-5601882048 N^5-1407467456 N^4+248136192 N^3-362165760 N^2 \nonumber\\
       &&-161782272 N+60134400~.
\end{eqnarray}
%----------------------------------------------------------------------------------------
Likewise, the new contributions to $a_{Qg}$ beyond the results given in \cite{Ablinger:2010ty}, 
are given by~:
%----------------------------------------------------------------------------------------
\begin{eqnarray}
%--------
a_{Qg}^{\sf b}(N) &=&
\textcolor{blue}{T_F C_A^2} \Biggl\{
%----
\frac{(8-N) }{12 N (N+1) (N+2)}S_1^5
%----
+\frac{Q_9 }{108 (N-1) N (N+1)^2 (N+2)^2}S_1^4
   \nonumber \\ &&
%----
+\Biggl[
\frac{Q_{20}}{54 (N-1) N^2 (N+1)^3 (N+2)^3}
+\frac{(88-39 N) }{18 N (N+1) (N+2)}S_2
\Biggr] S_1^3
   \nonumber \\ &&
%----
+\Biggl[
\frac{Q_{28}}{162 (N-1) N^3 (N+1)^4 (N+2)^4}
+\frac{Q_{11}}{18 (N-1) N (N+1)^2 (N+2)^2} S_2
   \nonumber \\ &&
+\frac{(8-5 N) }{3 N (N+1) (N+2)}S_3
-\frac{2 (N-72) }{3 N (N+1) (N+2)}S_{2,1}
+\frac{32 }{(N+1) (N+2)}S_{-2,1}
\Biggr] S_1^2
   \nonumber \\ &&
%----
+\Biggl[
\frac{(184-119 N) }{12 N (N+1) (N+2)}S_2^2
+\frac{Q_{19}}{54 (N-1) N^2 (N+1)^3 (N+2)^3} S_2
   \nonumber \\ &&
+\frac{Q_{32}}{81 (N-1) N^4 (N+1)^5 (N+2)^5}
+\frac{2 Q_5 }{27 (N-1) N (N+1)^2 (N+2)^2}S_3
   \nonumber \\ &&
+\frac{(53 N-8) }{6 N (N+1) (N+2)}S_4
+\frac{4 \big(95 N^3-787 N^2-2504 N-1310\big) }{9 N (N+1)^2 (N+2)^2}S_{2,1}
   \nonumber \\ &&
+\frac{4 (N+48) }{3 N (N+1) (N+2)}S_{3,1}
-\frac{32 \big(3 N^2+N-6\big) }{(N+1)^2 (N+2)^2}S_{-2,1}
+\frac{160 }{3 (N+1) (N+2)}S_{-2,2}
   \nonumber \\ &&
+\frac{64 }{(N+1) (N+2)}S_{-3,1}
-\frac{4 (5 N+48) }{3 N (N+1) (N+2)}S_{2,1,1}
-\frac{256 }{3 (N+1) (N+2)}S_{-2,1,1}
\Biggr] S_1
   \nonumber \\ &&
%----
+\frac{Q_{10}}{36 (N-1) N (N+1)^2 (N+2)^2} S_2^2
%----
-\frac{16 \big(3 N^2-23 N-20\big) }{3 (N-1) N (N+1)^2 (N+2)}S_{-2}^2
   \nonumber \\ &&
%----
+\frac{Q_{34}}{243 (N-1)^4 N^5 (N+1)^5 (N+2)^6}
%----
+\Biggl[
\frac{8 Q_7}{3 (N-1) N (N+1)^2 (N+2)^2}
   \nonumber \\ &&
-\frac{176 }{3 (N+1) (N+2)}S_1
\Biggr] S_{-4}
%----
+\Biggl[
-\frac{16 }{(N+1) (N+2)}S_1^2
+\frac{16 \big(3 N^2+N-6\big) }{(N+1)^2 (N+2)^2}S_1
   \nonumber \\ &&
-\frac{8 Q_{17}}{3 (N-1) N (N+1)^3 (N+2)^3}
-\frac{32 }{(N+1) (N+2)}S_2
\Biggr] S_{-3}
   \nonumber \\ &&
%----
+\frac{Q_{23}}{27 (N-1) N^2 (N+1)^3 (N+2)^3} S_3
%----
+\frac{Q_1 }{18 (N-1) N (N+1)^2 (N+2)^2}S_4
   \nonumber \\ &&
%----
+\frac{16 (11 N-16) }{3 N (N+1) (N+2)}S_5
%----
-\frac{296 }{3 (N+1) (N+2)}S_{-5}
%----
-\frac{2 Q_{15}}{27 N (N+1)^3 (N+2)^3} S_{2,1}
   \nonumber \\ &&
%----
+\Biggl[
-\frac{32 }{9 (N+1) (N+2)}S_1^3
+\frac{16 \big(3 N^2+N-6\big) }{3 (N+1)^2 (N+2)^2}S_1^2
+\Biggl[
      -\frac{16 Q_3}{3 (N+1)^3 (N+2)^3}
   \nonumber \\ &&
      -\frac{32 }{(N+1) (N+2)}S_2
\Biggr] S_1
+\frac{8 Q_{24}}{3 (N-1) N (N+1)^4 (N+2)^4}
+\frac{16 Q_4 }{(N-1) N (N+1)^2 (N+2)^2}S_2
   \nonumber \\ &&
-\frac{448 }{9 (N+1) (N+2)}S_3
+\frac{224 }{3 (N+1) (N+2)}S_{2,1}
\Biggr] S_{-2}
%----
+\frac{4 (N-8) }{3 N (N+1) (N+2)}S_{2,3}
   \nonumber \\ &&
%----
+\frac{176 }{3 (N+1) (N+2)}S_{2,-3}
%----
+\frac{4 \big(87 N^3-157 N^2-844 N-498\big) }{3 N (N+1)^2 (N+2)^2}S_{3,1}
%----
-\frac{56 }{3 (N+1) (N+2)}S_{4,1}
   \nonumber \\ &&
%----
+\frac{16 Q_3 }{(N+1)^3 (N+2)^3}S_{-2,1}
%----
+\Biggl[
\frac{Q_{31}}{162 (N-1)^2 N^3 (N+1)^4 (N+2)^4}
+\frac{(-39 N-584) }{9 N (N+1) (N+2)}S_3
   \nonumber \\ &&
+\frac{2 (9 N-8) }{3 N (N+1) (N+2)}S_{2,1}
+\frac{64 }{3 (N+1) (N+2)}S_{-2,1}
\Biggr] S_2 
%----
-\frac{80 \big(3 N^2+N-6\big) }{3 (N+1)^2 (N+2)^2}S_{-2,2}
   \nonumber \\ &&
%----
+\frac{48 }{(N+1) (N+2)}S_{-2,3}
%----
-\frac{32 \big(3 N^2+N-6\big) }{(N+1)^2 (N+2)^2}S_{-3,1}
%----
+\frac{400 }{3 (N+1) (N+2)}S_{-4,1}
   \nonumber \\ &&
%----
-\frac{8 \big(10 N^3-344 N^2-991 N-523\big) }{9 N (N+1)^2 (N+2)^2}S_{2,1,1}
%----
-\frac{224 }{3 (N+1) (N+2)}S_{2,1,-2}
   \nonumber \\ &&
%----
-\frac{4 (17 N-72) }{3 N (N+1) (N+2)}S_{2,2,1}
%----
-\frac{4 (23 N+8) }{3 N (N+1) (N+2)}S_{3,1,1}
%----
+\frac{128 \big(3 N^2+N-6\big) }{3 (N+1)^2 (N+2)^2}S_{-2,1,1}
   \nonumber \\ &&
%----
-\frac{64 }{(N+1) (N+2)}S_{-2,2,1}
%----
-\frac{80 }{(N+1) (N+2)}S_{-3,1,1}
%----
+\frac{8 (3 N+4) }{3 N (N+1) (N+2)}S_{2,1,1,1}
   \nonumber \\ &&
%----
+\frac{96 }{(N+1) (N+2)}S_{-2,1,1,1}
%----
+\Biggl[
\frac{(8-N) }{2 N (N+1) (N+2)}S_1^3
+\frac{Q_8 }{6 (N-1) N (N+1)^2 (N+2)^2}S_1^2
   \nonumber \\ &&
+\Biggl[
      \frac{Q_{21}}{6 (N-1) N^2 (N+1)^3 (N+2)^3}
      +\frac{(-3 N-8) }{2 N (N+1) (N+2)}S_2
\Biggr] S_1
   \nonumber \\ &&
+\frac{Q_{29}}{18 (N-1)^2 N^3 (N+1)^3 (N+2)^4}
+\Biggl[
      \frac{4 Q_4}{(N-1) N (N+1)^2 (N+2)^2}
   \nonumber \\ &&
      -\frac{8 }{(N+1) (N+2)}S_1
\Biggr] S_{-2}
+\frac{Q_2 }{6 (N-1) N (N+1)^2 (N+2)^2}S_2
+\frac{2 (3 N-8) }{N (N+1) (N+2)}S_3
   \nonumber \\ &&
-\frac{12 }{(N+1) (N+2)}S_{-3}
+\frac{16 }{N (N+1) (N+2)}S_{2,1}
+\frac{24 }{(N+1) (N+2)}S_{-2,1}
\Biggr] \zeta_2
   \nonumber \\ &&
%----
+\Biggl[
\frac{7 (N-8) }{3 N (N+1) (N+2)}S_1^2
-\frac{14 Q_6 }{3 (N-1) N (N+1)^2 (N+2)^2}S_1
   \nonumber \\ &&
+\frac{7 Q_{22}}{9 (N-1) N^2 (N+1)^2 (N+2)^3}
+\frac{7 (7 N+8) }{3 N (N+1) (N+2)}S_2
+\frac{56 }{3 (N+1) (N+2)}S_{-2}
\Biggr] \zeta_3
%----
\Biggr\}~,
\nonumber\\ 
\end{eqnarray}
%----------------------------------------------------------------------------------------
with the  polynomials $Q_i$ given by
%----------------------------------------------------------------------------------------
\begin{eqnarray}
Q_1&=&-3327 N^4-5641 N^3-5102 N^2-13268 N-7582 \\
Q_2&=&-51 N^4-361 N^3-434 N^2+196 N+290 \\
Q_3&=&3 N^4-6 N^3-21 N^2+24 N+52 \\
Q_4&=&3 N^4+N^3-24 N^2-60 N-40 \\
Q_5&=&5 N^4-427 N^3-4524 N^2-9344 N-5510 \\
Q_6&=&6 N^4-19 N^3-73 N^2-28 N-6 \\
Q_7&=&33 N^4+2 N^3-213 N^2-462 N-320 \\
Q_8&=&47 N^4-313 N^3-708 N^2+244 N+370 \\
Q_9&=&221 N^4-1849 N^3-3642 N^2+2212 N+2698 \\
Q_{10}&=&269 N^4-7345 N^3-13506 N^2+5188 N+6394 \\
Q_{11}&=&357 N^4-1381 N^3-4142 N^2-356 N+842 \\
%Q_{12}&=&10 N^5+209 N^4+293 N^3-76 N^2-36 N+80 \\
%Q_{13}&=&40 N^5-837 N^4-1175 N^3+968 N^2+604 N-80 \\
%Q_{14}&=&616 N^5+5199 N^4+7645 N^3+1392 N^2+748 N+3120 \\
Q_{15}&=&1504 N^5-8063 N^4-60746 N^3-111983 N^2-79376 N-21632 \\
%Q_{16}&=&N^6-11 N^5-44 N^4-67 N^3-97 N^2-78 N-64 \\
Q_{17}&=&9 N^6-N^5-19 N^4+87 N^3+488 N^2+940 N+608 \\
%Q_{18}&=&108 N^6-1394 N^5-4779 N^4-7811 N^3-11344 N^2-7148 N-6992 \\
Q_{19}&=&-1430 N^7+29061 N^6+168141 N^5+311889 N^4+262827 N^3+154488 N^2 \nonumber\\
       &&+113360 N+62400 \\
Q_{20}&=&-778 N^7+8151 N^6+39567 N^5+40819 N^4-14631 N^3-34136 N^2-12368 N \nonumber\\
       &&+1600 \\
Q_{21}&=&-122 N^7+1331 N^6+7459 N^5+10911 N^4+4621 N^3+648 N^2+1776 N+1600 \\
Q_{22}&=&20 N^7+18 N^6-333 N^5-1450 N^4-3273 N^3-4570 N^2-3692 N-1480 \\
Q_{23}&=&-1080 N^8+1898 N^7+22443 N^6+92307 N^5+267403 N^4+421473 N^3 \nonumber\\
       &&+361900 N^2+222376 N+81520 \\
Q_{24}&=&3 N^8+21 N^7+174 N^6+700 N^5+1506 N^4+1216 N^3-1676 N^2-4024 N-2144 \\
%Q_{25}&=&770 N^8+19184 N^7+58735 N^6+43361 N^5-33518 N^4-57052 N^3-34168 N^2 \nonumber\\
%       &&-6528 N-2880 \\
%Q_{26}&=&8 N^{10}-123 N^9-779 N^8-1330 N^7+549 N^6+3768 N^5+2110 N^4-423 N^3 \nonumber\\
%       &&+536 N^2-1244 N-912 \\
%Q_{27}&=&576 N^{10}-9342 N^9-61190 N^8-123779 N^7+7766 N^6+328327 N^5+232310 N^4 \nonumber\\
%       &&-58484 N^3+1096 N^2-93216 N-68544 \\
Q_{28}&=&2080 N^{10}-155867 N^9-992144 N^8-2266725 N^7-2100345 N^6+59166 N^5 \nonumber\\
       &&+1939307 N^4+1900784 N^3+806384 N^2+252096 N+57600 \\
Q_{29}&=&184 N^{11}+176 N^{10}-4108 N^9-16762 N^8-25657 N^7-15063 N^6-883 N^5 \nonumber\\
       &&+14485 N^4+37996 N^3+13360 N^2-32048 N-17040 \\
%Q_{30}&=&3160 N^{11}+44348 N^{10}+194250 N^9+381049 N^8+309309 N^7-74100 N^6 \nonumber\\
%       &&-359068 N^5-255860 N^4+1008 N^3+74240 N^2+43392 N+11520 \\
Q_{31}&=&6624 N^{12}+5292 N^{11}-210971 N^{10}-1104257 N^9-2480453 N^8-2265264 N^7 \nonumber\\
       &&+636087 N^6+2871225 N^5+2408885 N^4+828944 N^3-1329328 N^2 \nonumber\\
       &&-1961664 N-671040 \\
Q_{32}&=&15604 N^{13}+361847 N^{12}+2453891 N^{11}+8204366 N^{10}+15666936 N^9 \nonumber\\
       &&+16766294 N^8+5755934 N^7-9519761 N^6-13953239 N^5-6072896 N^4 \nonumber\\
       &&+1787904 N^3+3241984 N^2+1353984 N+230400 \\
%Q_{33}&=&230 N^{18}-9103 N^{17}-98422 N^{16}-326320 N^{15}-227272 N^{14}+803071 N^{13} \nonumber\\
%       &&+1398595 N^{12}+11553 N^{11}-1472228 N^{10}-1374447 N^9+712249 N^8 \nonumber\\
%       &&+2701565 N^7-493236 N^6-2546615 N^5+785364 N^4+1166104 N^3+155904 N^2 \nonumber\\
%       &&-340272 N-146880 \\
Q_{34}&=&6208 N^{19}-86928 N^{18}-1344972 N^{17}-6002889 N^{16}-9808011 N^{15}+4340125 N^{14} \nonumber\\
       &&+32811393 N^{13}+24313093 N^{12}-30513058 N^{11}-46961276 N^{10}+3785621 N^9 \nonumber\\
       &&+36663986 N^8+1686347 N^7-40115539 N^6-14945624 N^5+25303412 N^4 \nonumber\\
       &&+16493728 N^3-1302672 N^2-6643584 N-2376000~.
\end{eqnarray}
%---------------------------------------------------------------------------------------------------------------
These quantities will be used in the later calculation of the full massive OMEs.
%---------------------------------------------------------------------------------------------------------------
\section{Complete Wilson Coefficients}
%---------------------------------------------------------------------------------------------------------------

\noindent
After the first two massive OMEs, $L_{qg,Q}^{(3)}$ and $L_{qq,Q}^{(3),\rm PS}$, at 3-loop order
were calculated in \cite{Ablinger:2010ty}, during the last months we computed four other OMEs and 
associated massive Wilson 
coefficients in the asymptotic region $Q^2 \gg m^2$. These are the non-singlet OME 
$A_{qq,Q}^{(3), \rm NS}$, that of transversity $A_{qq,Q}^{(3), \rm NS, TR}$,
$A_{gq,Q}^{(3)}$, and very recently also the pure singlet $A_{gq,Q}^{(3),\sf PS}$. These matrix elements 
contain topologies up to Benz-graphs with respective local operator insertions. We used {\tt Reduze2} 
\cite{Studerus:2009ye,vonManteuffel:2012np}
to reduce the diagrams to master integrals applying the integration-by-parts relations for Feynman 
diagrams containing local operator insertions. In the first three cases the master integrals could be 
calculated using hypergeometric function techniques and Mellin-Barnes 
\cite{Czakon:2005rk,Smirnov:2009up}
representations to map the integrals into nested
finite and infinite sums, which were then solved using the summation technologies of 
\cite{Schneider08JSC, 
      Schneider05AC,
      Schneider06JDEA, 
      Schneider10AC, 
      Schneider10PW, 
      Schneider06SL,
      Schneider07Hab, 
      Ablinger:2010pb, 
      Blumlein:2012hg,
      Ablinger:2010ha, 
      Schneider2013,
      Ablinger:2013cf,
      Ablinger:2013hcp}.
For 
$A_{qq,Q}^{(3), \rm NS}, A_{qq,Q}^{(3), \rm NS, TR}$ and $A_{gq,Q}^{(3)}$ the results can be represented 
using harmonic sums 
only.

As an example we show the constant part of the unrenormalized OME for transversity $a_{qq}^{\rm NS,TR(3)}$
for even and odd values of $N$~:
%---------------------------------------------------------------------------------------------------------------
\begin{eqnarray}
a_{qq}^{\rm NS,TR(3)}(N) &=& 
\textcolor{blue}{C_F^2 T_F}
\Biggl\{
\frac{128}{27} S_2 S_1^3
+\Biggl[\frac{64}{3} S_3
-\frac{128}{9} S_{2,1}
-\frac{256}{9} S_{-2,1}
-\frac{16}{9 N}
-\frac{32 (-1)^N}{9 N (N+1)}
\Biggr] S_1^2
\nonumber\\ &&
%---
+\Biggl[
-\frac{64}{9} S_2^2
+\frac{7168 S_2}{81}
+\frac{32 (-1)^N (13 N+7)}{27 N (N+1)^2}
-\frac{2560 S_3}{27}
+\frac{704 S_4}{9}
-\frac{320}{9} S_{3,1}
\nonumber\\ &&
-\frac{2560}{27} S_{-2,1}
-\frac{256}{9} S_{-2,2}
+\frac{64}{3} S_{2,1,1}
+\frac{1024}{9} S_{-2,1,1}
\nonumber\\ &&
+\frac{8 \big(769 N^4+1547 N^3+787 N^2-15 N-12\big)}{27 N^2 (N+1)^2}
\Biggr] S_1
%---
-\frac{496}{27} S_2^2
\nonumber\\ &&
-\frac{16 (-1)^N \big(133 N^4+188 N^3+46 N^2-45 N-18\big)}{81 N^3 (N+1)^3}
\nonumber\\ &&
-\frac{2 \big(6327 N^6+18981 N^5+18457 N^4+5687 N^3-260 N^2+144 N+144\big)}{81 N^3 (N+1)^3}
%---
\nonumber\\ &&
+ \Biggl[
16-\frac{64}{3} S_1
\Biggr] B_4
%---
+ 
\Biggl[
\frac{256}{9} S_1
-\frac{1280}{27}\Biggr] S_{-4}
%---
+\Biggl[96 S_1-72\Biggr] \zeta_4
%---
+\Biggl[
\frac{128}{9} S_1^2-\frac{1280}{27} S_1
\nonumber\\ &&
+\frac{128}{9} S_2 +\frac{7168}{81}
\Biggr] S_{-3}
%---
+\frac{10408}{81} S_3
-\frac{2992}{27} S_4
+\frac{512}{9} S_5
+\frac{256}{9} S_{-5}
%---
+
\Biggl[
\frac{256}{27} S_1^3
\nonumber\\ &&
+\frac{14336}{81} S_1
-\frac{1280}{27} S_2 +\frac{512}{27} S_3
-\frac{512}{9} S_{2,1}-\frac{64}{9 N (N+1)}
\Biggr] S_{-2}
%---
+\frac{112}{9} S_{2,1}
+\frac{256}{9} S_{2,3}
\nonumber\\ &&
-\frac{512}{9} S_{2,-3}
+\frac{1424}{27} S_{3,1}
-\frac{512}{9} S_{4,1}
-\frac{14336}{81} S_{-2,1}
%---
+\Biggl[
-\frac{16 \big(169 N^2+169 N+6\big)}{27 N (N+1)}
\nonumber\\ &&
+\frac{256 S_3}{27}
+\frac{256}{3} S_{-2,1}
-\frac{32 (-1)^N}{9 N (N+1)}\Biggr] S_2
-\frac{1280}{27} S_{-2,2}
+\frac{512}{9} S_{-2,3}
-16 S_{2,1,1}
+\frac{512}{9} S_{2,1,-2}
\nonumber\\ &&
+\frac{256}{9} S_{3,1,1}
+\frac{5120}{27} S_{-2,1,1}
+\frac{512}{9} S_{-2,2,1}
-\frac{2048}{9} S_{-2,1,1,1}
\nonumber\\ &&
%---
+\Biggl[-\frac{2 \big(45 N^2+45 N-4\big)}{3 N (N+1)}
+\frac{64}{3} S_{-2} S_1
-8 S_2
+ \Biggl[\frac{32}{3} S_2+40\Biggr] S_1
+\frac{32}{3} S_3
+\frac{32}{3} S_{-3}
\nonumber\\ &&
-\frac{64}{3} S_{-2,1}
-\frac{8 (-1)^N}{3 N (N+1)}\Biggr]\zeta_2
%---
+\Biggl[-\frac{1208}{9} S_1
-\frac{64}{3} S_2
+\frac{350}{3}\Biggr] \zeta_3
\Biggr\} 
%-------------------------------------------------------------------
\nonumber\\ &&
+
\textcolor{blue}{C_F T_F^2}
\Biggl\{
\frac{8 \big(157 N^4+314 N^3+277 N^2-24 N-72\big)}{243 N^2 (N+1)^2}
-\frac{19424}{729} S_1 +\frac{1856}{81} S_2
-\frac{640}{81} S_3
\nonumber\\ &&
+\frac{128}{27} S_4
%---
+ \textcolor{blue}{N_F} 
\Biggl[
\frac{32 \big(308 N^4+616 N^3+323 N^2-3 N-9\big)}{243 N^2 (N+1)^2}
-\frac{55552}{729} S_1 +\frac{640}{27} S_2
\nonumber\\ &&
-\frac{320}{81} S_3
+\frac{64}{27} S_4 \Biggr]
%---
+\Biggl[-\frac{320}{27} S_1
+\frac{64}{9} S_2
%---
+ \textcolor{blue}{N_F} \Biggl[
-\frac{160}{27} S_1
+\frac{32}{9} S_2 
+\frac{16}{9}\Biggr]
+\frac{32}{9}
\Biggr] \zeta_2
\nonumber\\ &&
%---
+\Biggl[
-\frac{1024}{27} S_1
%---
+ \textcolor{blue}{N_F} \Biggl[
\frac{448}{27} S_1
-\frac{112}{9}\Biggr]
+\frac{256}{9}\Biggr] \zeta_3 \Biggr\} 
\nonumber\\
&& + \textcolor{blue}{C_A C_F T_F} 
\Biggl\{
-\frac{64}{27} S_2 S_1^3
+\Biggl[
\frac{4 (3 N+2)}{9 N (N+1)}
-\frac{80}{9} S_3
+\frac{128}{9} S_{2,1}
+\frac{128}{9} S_{-2,1}
+\frac{16 (-1)^N}{9 N (N+1)}
\Biggr] S_1^2
%---
\nonumber\\ &&
+\Biggl[
\frac{112}{9} S_2^2
-\frac{16 (N-2) (2 N+3)}{9 (N+1) (N+2)} S_2
-\frac{16 (-1)^N (13 N+7)}{27 N (N+1)^2}
\nonumber\\ &&
+\frac{4 \big(6197 N^3+18591 N^2+15850 N+4320\big)}{729 N (N+1) (N+2)}
+\frac{320}{9} S_3
-\frac{208}{9} S_4
-8 S_{2,1}
+\frac{64}{3} S_{3,1}
\nonumber\\ &&
+\frac{1280}{27} S_{-2,1}
+\frac{128}{9} S_{-2,2}
-32 S_{2,1,1}
-\frac{512}{9} S_{-2,1,1}
\Biggr] S_1
%---
-\frac{20}{3} S_2^2
\nonumber\\ &&
+\frac{8 (-1)^N \big(133 N^4+188 N^3+46 N^2-45 N-18\big)}{81 N^3 (N+1)^3}
\nonumber\\ &&
+\frac{-1013 N^6-3039 N^5-5751 N^4-2981 N^3+1752 N^2+1872 N+432}{243 N^3 (N+1)^3}
%---
\nonumber\\ &&
+ \Biggl[72-96 S_1\Biggr] \zeta_4
+ \Biggl[\frac{640}{27}
- \frac{128}{9} S_1\Biggr] S_{-4} 
+ \Biggl[
\frac{32}{3} S_1-8\Biggr] B_4
+ \Biggl[
-\frac{64}{9} S_1^2
+\frac{640}{27} S_1
\nonumber\\ &&
-\frac{64}{9} S_2
-\frac{3584}{81}\Biggr] S_{-3}
-\frac{8 \big(27 N^3+560 N^2+1365 N+778\big)}{81 (N+1) (N+2)} S_3
+\frac{1244}{27} S_4
-\frac{224}{9} S_5
\nonumber\\ &&
-\frac{128}{9} S_{-5}
-\frac{32 \big(3 N^3+7 N^2+7 N+6\big)}{9 (N+1) (N+2)} S_{2,1}
%---
+\Biggl[-\frac{128}{27} S_1^3-\frac{7168}{81} S_1
\nonumber\\ &&
+\frac{640}{27} S_2
-\frac{256}{27} S_3
+\frac{256}{9} S_{2,1}
+\frac{32}{9 N (N+1)}\Biggr] S_{-2}
%---
-\frac{128}{3} S_{2,3}
+\frac{256}{9} S_{2,-3}
-\frac{1352}{27} S_{3,1}
\nonumber\\ &&
+\frac{256}{9} S_{4,1}
%---
+ \Biggl[-\frac{4 \big(364 N^3+1227 N^2+872 N+36\big)}{81 N (N+1) (N+2)}
+\frac{496}{27} S_3
-\frac{64}{3} S_{2,1}
\nonumber\\ &&
-\frac{128}{3} S_{-2,1}
+\frac{16 (-1)^N}{9 N (N+1)}\Biggr] S_2
+\frac{7168}{81} S_{-2,1}
+\frac{640}{27} S_{-2,2}
-\frac{256}{9} S_{-2,3}
+24 S_{2,1,1}
\nonumber\\ &&
-\frac{256}{9} S_{2,1,-2}
+\frac{64}{3} S_{2,2,1}
-\frac{256}{9} S_{3,1,1}
-\frac{2560}{27} S_{-2,1,1}
-\frac{256}{9} S_{-2,2,1}
+\frac{224}{9} S_{2,1,1,1}
\nonumber\\ &&
+\frac{1024}{9} S_{-2,1,1,1}
%---
+\Biggl[
\frac{2 \big(35 N^2+35 N-6\big)}{9 N (N+1)}
-\frac{32}{3} S_{-2} S_1
-\frac{16}{27} S_1
-\frac{88}{9} S_2
-\frac{16}{3} S_3
\nonumber\\ &&
-\frac{16}{3} S_{-3}
+\frac{32}{3} S_{-2,1}
+\frac{4 (-1)^N}{3 N (N+1)}\Biggr] \zeta_2
%---
+\Biggl[
-16 S_1^2
+\frac{2548 S_1}{27}
\nonumber\\ &&
+\frac{2 \big(108 N^3-239 N^2-1137 N-646\big)}{9 (N+1) (N+2)}
+16 S_2\Biggr] \zeta_3
\Biggr\}~,
\nonumber
\end{eqnarray}
%---------------------------------------------------------------------------------------------------------------
with $B_4$  
%---------------------------------------------------------------------------------------------------------------
\begin{eqnarray}
B_4 = - 4 \zeta_2 \ln^2(2) + \frac{2}{3} \ln^4(2) - \frac{13}{2} \zeta_4 + 16 \Li_4\left(\frac{1}{2}\right)~.
\end{eqnarray}
%---------------------------------------------------------------------------------------------------------------
It is represented by harmonic sums up to {\sf w = 5}. The logarithmic contributions and other pieces
of the constant term stemming from lower order quantities are given in \cite{LOG}. The renormalized
expression both in $N$ and $x$-space is presented in \cite{NS1}. The analytic continuation to 
complex values of $N$ is obtained using the relations being given in 
Refs.~\cite{Blumlein:2009ta,ANC}.

Very recently also the massive OME of the pure singlet case has been calculated \cite{PS}. Here 
the master integrals
are somewhat more demanding than for $A_{gq}$ and $A_{qq}^{\sf NS, TR}$ and they were partly solved 
using
differential and difference equations applying the packages \CSsigma, \EvalMS, \SumProd
and {\tt HarmonicSums}. Here the structure of the Wilson coefficient contains 
a series of 
generalized harmonic sums as a fully inclusive quantity in QCD. They are of the type
%---------------------------------------------------------------------------------------------------------------
\begin{eqnarray}
S_{2,1}(2,1;N),~~~S_{1,1,2}\left(2,\frac{1}{2},1;N\right),~~~S_{3}(2;N),~~~~~\text{etc.}
\end{eqnarray}
%---------------------------------------------------------------------------------------------------------------
These sums may individually diverge as $N \rightarrow \infty$. However, the asymptotic expansion of the
complete expression is well behaved. In the representation in $x$-space, generalized harmonic 
polylogarithms emerge. For QCD corrections being related to deep-inelastic scattering quantities
of this kind are observed for the first time.

%%%%%%%%%%%%%%%%%%%%%%%%%%%%%%%%%%%%%%%%%%%%%%%%%%%%%%%%%%%%%%%%%%%%%%%%%%%%%%%%%%%%%%%%%%%%%%%%%%
\section{Graphs with two massive quark lines of equal masses}
%%%%%%%%%%%%%%%%%%%%%%%%%%%%%%%%%%%%%%%%%%%%%%%%%%%%%%%%%%%%%%%%%%%%%%%%%%%%%%%%%%%%%%%%%%%%%%%%%%

\noindent
Starting from 3-loop order graphs with two distinct internal massive lines
occur in the calculation of the massive operator matrix elements. The
corresponding contributions to the operator matrix elements $A_{gg}$
and $A_{gq}$ are characterized by the color factors $T_F^2 C_A (C_F)$ without 
additional factors $Nf$.  The challenge in computing these
diagrams  derives from the fact that identifying hypergeometric series
directly, as used in earlier 3-loop calculations \cite{Ablinger:2010ty,
Ablinger:2012qm, Blumlein:2012vq}, leads to divergent sums. In fact the degree of
divergence even grows linearly with $N$ due to factors
%--------------------------------------------------------------------------------
\begin{align}
  B(N+i+\alpha, -i+\beta) 
  =
  \frac{\Gamma(N+i+\alpha)\Gamma(-i+\beta)}{\Gamma(N+\alpha+\beta)}~,
\end{align}
%--------------------------------------------------------------------------------
where $N$ is the Mellin variable, $i$ the summation index of an
infinite sum, and $\alpha, \beta$ are independent of $N$ and $i$. To
avoid this source of divergence, a Mellin-Barnes representation is
introduced and the Beta-function of the above type is kept in the form
of a Feynman parameter integral. As a result, the divergent pattern
can be removed by observing that the contour of the Mellin-Barnes
integral either must be closed to the right {\it or} to the
left, depending on the value of the remaining Feynman parameter.  Due
to this distinction the remaining integral does not represent
a Beta-function anymore, but will be performed in the space of certain
iterated integrals at a later stage. The sum of residues is simplified
using symbolic summation technologies \cite{Schneider08JSC,
Schneider05AC, Schneider06JDEA, Schneider10AC, Schneider10PW,
Schneider06SL, Schneider07Hab, Ablinger:2010pb, Ablinger:2010ha,
Schneider2013, Blumlein:2012hg}. 

In order to perform the last integral in $x$, say, a generating
function for the Mellin moments is introduced with
%--------------------------------------------------------------------------------
\begin{align}
 \sum_{N=0}^{\infty}
 (\kappa R(x))^N
 =
 \frac{1}{1-\kappa R(x)}
\comma
\end{align}
%--------------------------------------------------------------------------------
where $R(x)\in \{1/(1+x^2),x^2/(1+x^2)\}$. This introduces cyclotomic
letters weighted by the tracing parameter $\kappa$. The resulting
expression involves cyclotomic harmonic polylogarithms (HPLs) which depend on the 
variable $\kappa$. In order to extract the Mellin-space expression the
$N$th Taylor coefficient has to be calculated, which is possible using
the packages \HSums{} \citeHSums{} and \CSsigma{} \citeCSsigma{}. The
resulting multi-sum expressions are simplified using the package
\EvalMS{} \citeEvalMS{} and expressed in a basis of indefinite
(nested) sums. 

\begin{figure}[H]
  \centering
  \includegraphics[width=0.2\textwidth]{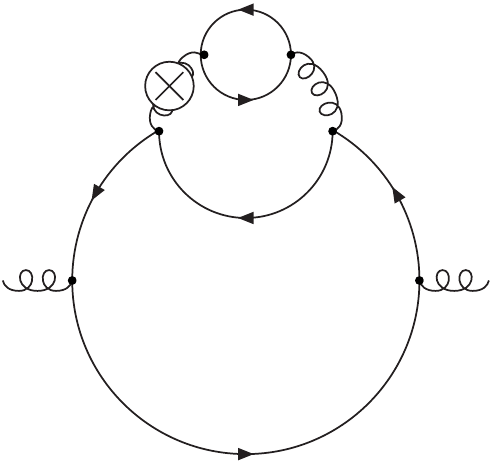}
  \caption{\small A digram with two massive quark cycles contributing to $A_{gg}$.}
  \label{fig:D560}
\end{figure}

As a proof of principle we calculated all scalar graphs which
correspond to the $T_F^2$-contributions to $A_{gg}$. Also the
calculation of the full $T_F^2$ contributions will be finished soon.
One of the QCD-graphs is shown in Figure~\ref{fig:D560}.
%--------------------------------------------------------------------------
\restylefloat{figure}
%--------------------------------------------------------------------------
\noindent
The $N$-space result has the following form:
%--------------------------------------------------------------------------
\begin{align}
\label{eq:D560Res}
    I_{560}
    ={}&
     \frac{2\overline{P}_4}{3 N (N+1)^2 (N+2) (2 N-5) (2 N-3) (2 N-1)}\frac{1}{4^N}\binom{2N}{N}
%\N\\\times
     \Bigg[
        \sum_{j=1}^N \frac{4^{j} S_1\big(j\big)}{\binom{2 j}{j} j^2}
       -\sum_{j=1}^N \frac{4^{j}}{\binom{2 j}{j} j^3}
       -7 \zeta_3
     \Bigg]
\N\\ &
    +\frac{N^2+N+2}{27 (N-1) N^2 (N+1)}
     \Big[-144 S_{2,1}
          -36 \zeta_2 S_1-4 S_1^3+36 S_2 S_1 + 88 S_3
\N\\ &
+312 \zeta_3\Big]
    -\frac{4 \overline{P}_2}{6075 (N-2)^2 (N-1)^4 N^5 (N+1)^4 (N+2) (2 N-5) (2 N-3) (2 N-1)}
\N\\&
    -\frac{64 \big(N^2+N+2\big)}{9 \ep^3 (N-1) N^2 (N+1)}
    +\frac{1}{\ep^2}\Big\{
      -\frac{32 \overline{P}_6}{27 (N-1)^2 N^3 (N+1)^2 (N+2)}
      -\frac{32 \big(N^2+N+2\big)}{9 (N-1) N^2 (N+1)} S_1
     \Big\}
\N\\&
    +\frac{1}{\ep}\Big\{
      -\frac{8 \overline{P}_7}{405 (N-2) (N-1)^3 N^4 (N+1)^3 (N+2)}
      -\frac{16 \overline{P}_6}{27 (N-1)^2 N^3 (N+1)^2 (N+2)} S_1
\N\\&
      -\frac{8 \big(N^2+N+2\big) \big(S_1^2-3 S_2 + 3 \zeta_2\big)}{9 (N-1) N^2 (N+1)}
      -\frac{8 \big(55 N^3+235 N^2-52 N+20\big)}{15 (N-2) (N-1) N (N+1)^2 (N+2)}
     \Big\}
\N\\&
    -\frac{4 \overline{P}_5 S_1}{81 (N-1)^3 N^4 (N+1)^3 (N+2) (2 N-5) (2 N-3) (2 N-1)}
\N\\&
    -\frac{4 \overline{P}_6 \big(S_1^2-3 S_2+3 \zeta_2\big)}{27 (N-1)^2 N^3 (N+1)^2 (N+2)}
    -\frac{4 \overline{P}_1}{225 (N-2)^2 (N-1)^2 N^2 (N+1)^3 (N+2)}~.
\end{align}
%--------------------------------------------------------------------------
Here the functions $\overline{P}_i$ are polynomials in $N$ up to degree $d = 17$ and we used
the shorthand notation $S_{\vec{a}}(N) \equiv S_{\vec{a}}$ for the harmonic sums.
Besides the well-known harmonic sums the above diagram depends on the new structure
%--------------------------------------------------------------------------
\begin{align}
\label{eq:invbinsumcomb}
  \frac{1}{4^N}
  \binom{2N}{N}
  \Biggl[
     \sum_{j=1}^N 
     \frac{4^{j} S_{1}(j)}
     {\binom{2j}{j} j^2}
   - \sum _{j=1}^N 
     \frac{4^{j} }
     {\binom{2j}{j} j^3}
   - 7 \zeta_3
   \Biggr] \comma
\end{align}
%--------------------------------------------------------------------------
which involves binomially weighted harmonic sums within finite sums.\footnote{Infinite binomial and
inverse binomial sums have been considered in Refs.~\cite{Fleischer:1998nb,Kalmykov:2000qe, 
Davydychev:2003mv,Weinzierl:2004bn}.} These objects cannot be represented in terms
of (generalized) harmonic sums or (generalized) cyclotomic sums. In all scalar
diagrams, and all considered QCD diagrams contributing to the color factors $T_F^2 C_A (C_F)$
these sums occur in the same combination, which is hence a property of the corresponding Feynman
diagrams. In some terms denominators occur, which introduce poles at points
$N=\frac{1}{2}, \frac{3}{2}, 2, \frac{5}{2}$. However, the rightmost singularity expected for
these diagrams is $N = 1$. Interestingly, these poles can be shown to be removable by expanding 
(\ref{eq:D560Res}) in a Laurent series around these points.

%%%%%%%%%%%%%%%%%%%%%%%%%%%%%%%%%%%%%%%%%%%%%%%%%%%%%%%%%%%%%%%%%%%%%%%%%%%%%%%%%%%%%%%%%%%%%%%%%%
\section{Massive quark production in charged current DIS at 2-loop order}
%%%%%%%%%%%%%%%%%%%%%%%%%%%%%%%%%%%%%%%%%%%%%%%%%%%%%%%%%%%%%%%%%%%%%%%%%%%%%%%%%%%%%%%%%%%%%%%%%%

\noindent
The $O(\alpha_s)$ corrections to heavy flavor production in charged current deep-inelastic
scattering have been calculated in \cite{Gottschalk:1980rv,Kretzer99,Blumlein:2011zu}. Here the
$O(\alpha_s^2)$ corrections are presented in the asymptotic region $Q^2 \gg m^2$ \cite{BHP13}, 
comparing to an earlier calculation in Ref.~\cite{Buza:1997mg}. This process is particularly 
important because of its sensitivity to the sea quark densities $\bar{s}(x,Q^2), \bar{d}(x,Q^2)$ and
$\bar{u}(x,Q^2)$. Furthermore the asymptotic representation is fully justified since the 
corresponding data are measured mostly at high virtualities $Q^2~\gsim~100~\GeV^2$.\footnote{For 
a discussion of the accuracy of the asymptotic representation see \cite{Buza:1995ie}.}

The charged current cross sections for deep inelastic lepton-nucleon scattering is commonly
parameterized in three structure functions $F_1$, $F_2$, $F_3$~:
%---------------------------------------------------------------------
\begin{align}
\label{eq:XSa}
 \frac{d\sigma^{\nu(\bar{\nu})}}{dx dy}
 ={}&
 \frac{G_F^2 s}{4 \pi}
 \left\{
      (1+(1-y)^2) F_2^{W^{\pm}}
  - y^2 F_L^{W^{\pm}}
  \pm (1-(1-y)^2) x F_3^{W^{\pm}}
 \right\}
\comma
\\
\label{eq:XS}
 \frac{d\sigma^{e^-(e^+)}}{dx dy}
 ={}&
 \frac{G_F^2 s}{4 \pi}
 \left\{
      (1+(1-y)^2) F_2^{W^{\mp}}
  - y^2 F_L^{W^{\mp}}
  \pm (1-(1-y)^2) x F_3^{W^{\mp}}
 \right\}
\period
\end{align}
%---------------------------------------------------------------------

The expressions of the heavy flavor Wilson coefficients in the asymptotic region are
constructed in terms of light flavor Wilson coefficients and massive
operator matrix elements (OMEs).  This is achieved by exploiting the
process independence of the PDFs and OMEs and constructing the
4-flavor expressions in the variable flavor number scheme in
\cite{Buza:1996wv,Bierenbaum:2009mv}. By matching them back onto the 3-flavor
scheme one finds the factorization formulae in the asymptotic region~:
%---------------------------------------------------------------------
\begin{align}
  L_{i,q}^{W^+\pm W^-,\text{NS},(2)}
  =\;&
  \delta_{i,2}
  A_{qq,Q}^{\text{NS},(2)}
  + C_{i,q}^{W^+\pm W^-,\text{NS},(2)}(n_f+1)
  - C_{i,q}^{W^+\pm W^-,\text{NS},(2)} (n_f)
\comma
\N\\
  H_{i,q}^{W^+\pm W^-,\text{NS},(2)}
  =\;&
  \delta_{i,2} 
  A_{qq,Q}^{\text{NS},(2)}
  +C_{i,q}^{W^+\pm W^-,\text{NS},(2)}(n_f+1)
\comma
\N\\ 
  L_{i,q}^{W,\text{PS},(2)}
  =\;&
    C_{i,q}^{W,\text{PS},(2)}(n_f+1)
  - C_{i,q}^{W,\text{PS},(2)}(n_f)
  =0
\comma
\N\\
  H_{i,q}^{W,\text{PS},(2)}
  =\;&
  \frac{1}{2}
  \delta_{i,2} 
  A_{Qq}^{\text{PS},(2)}
  +C_{i,q}^{W,\text{PS},(2)}(n_f+1)
\comma
\N\\
  L_{i,g}^{W,(2)}
  =\;&
  A_{gg,Q}^{(1)}
  C_{i,g}^{W,(1)}(n_f+1)
  +C_{i,g}^{W,(2)}(n_f+1)
  -C_{i,g}^{W,(2)}(n_f)
\comma
\N\\
  H_{i,g}^{W,(2)}
  =\;&
  A_{gg,Q}^{(1)}
  C_{i,g}^{W,(1)}(n_f+1)
  +C_{i,g}^{W,(2)}(n_f+1)
\N\\&
  +\frac{1}{2} 
  \left(
    \delta_{i,2}
    A_{Qg}^{(2)}
   +A_{Qg}^{(1)}
    C_{i,q}^{W^++ W^-,\text{NS},(1)}(n_f+1)
  \right)
\comma
\N\\
  L_{3,q}^{W^+\pm W^-,\text{NS},(2)}
  =\;&
  A_{qq,Q}^{\text{NS},(2)}
  +C_{3,q}^{W^+\pm W^-,\text{NS},(2)}(n_f+1)
  -C_{3,q}^{W^+\pm W^-,\text{NS},(2)}(n_f)
\comma
\N\\
  H_{3,q}^{W^+\pm W^-,\text{NS},(2)}
  =\;&
  A_{qq,Q}^{\text{NS},(2)}
  +C_{3,q}^{W^+\pm W^-,\text{NS},(2)}(n_f+1)
\comma
\N\\
  H_{3,q}^{W,\text{PS},(2)}
  =\;&
  -\frac{1}{2} A_{Qq}^{\text{PS},(2)}
\comma
\N\\
  H_{3,g}^{W,(2)}
  =\;&
  \frac{1}{2} 
  \left(
    -A_{Qg}^{(2)}
    -A_{Qg}^{(1)}
     C_{3,q}^{W^++ W^-,\text{NS},(1)}(n_f+1)
   \right)
\period
\label{eq:WC}
\end{align}
%---------------------------------------------------------------------
Here $A_{ij}^{(k)},~k =1,2$ denote the massive OMEs \cite{Buza:1995ie,Bierenbaum:2007dm,Bierenbaum:2007qe,
Bierenbaum:2008yu,Bierenbaum:2009zt} and $C_{l,m}$ the massless Wilson coefficients 
\cite{Zijlstra:1991qc,vanNeerven:1991nn,Zijlstra:1992kj,Zijlstra:1992qd,Moch:1999eb}
up to 2-loop order. Note that the $O(\alpha_s^2)$ corrections contain, besides a single heavy quark
excitation also contributions due to heavy quark pair-production.

The Wilson coefficients (\ref{eq:WC}) correct and complete an earlier derivation of the
heavy flavor Wilson coefficients \cite{Buza:1997mg}; for details see Ref.~\cite{BHP13}. 
The difference lies in factors $(-1)$ in the expressions for $H_{3,q}^{W,\text{PS},(2)}$ 
and $H_{3,g}^{W,(2)}$. The correctness of the present result was checked by an explicit 
calculation of the leading logarithmic parts using the same idea as in \cite{Altarelli:1977zs},
referring to the ladder graph contributions in physical gauge. Furthermore the construction of 
$H_{3,g}^{W,(1)}$ in the same way delivers a minus sign that can be reproduced by the asymptotic
expansion of the exact 1-loop result \cite{Blumlein:2011zu}. We also calculated the terms to $O(\alpha_s^2)$
having been left out in  Ref.~\cite{Buza:1997mg}. 

The representation of the Wilson coefficients (\ref{eq:WC}) both in Mellin-$N$ and $x$-space have 
been derived \cite{BHP13} using the package \HSums{} \citeHSums{}. For the use in phenomenological applications 
we implemented both these expressions into {\tt FORTRAN}-programs, which are available on request.

\begin{figure}[H]
 \centering
 \includegraphics[width=.49\textwidth]{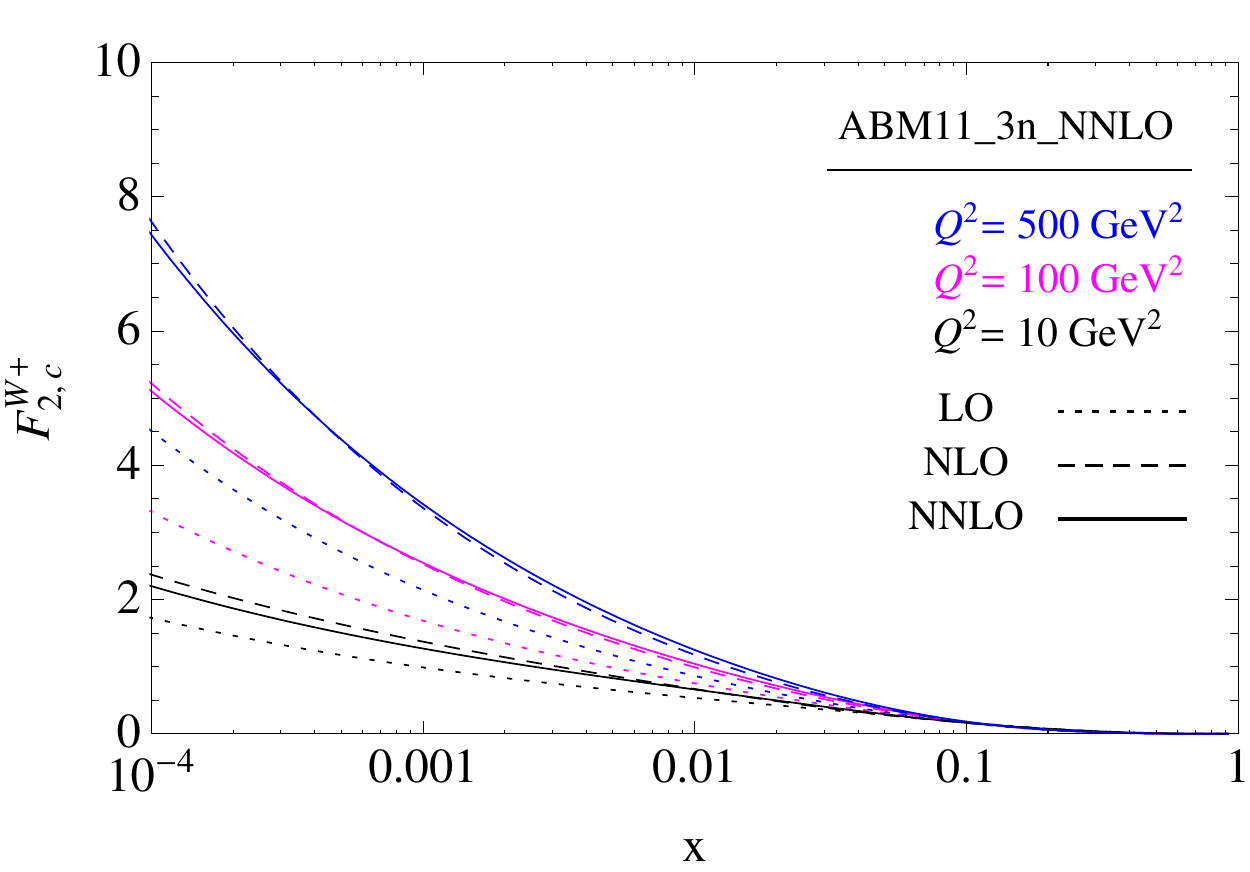}
 \includegraphics[width=.49\textwidth]{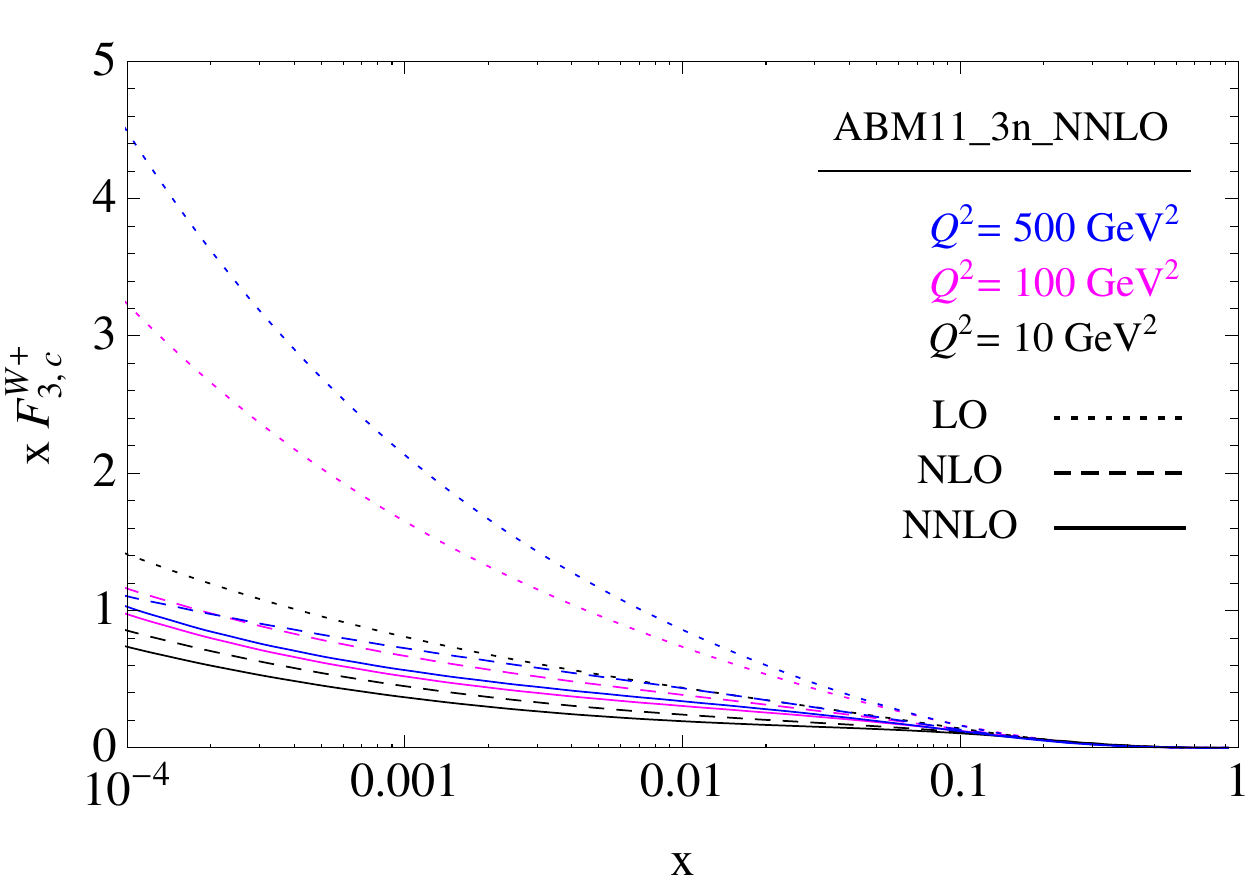}
 \caption{\small Charm contributions to the structure functions $F_2$ and
	  $xF_3$ of deep-inelastic scattering via $W^+$-exchange at
          LO, NLO, NNLO.}
 \label{fig:F2cWp}
\end{figure}

In Figures~\ref{fig:F2cWp} numerical illustrations of the charm quark corrections to $F_2(x,Q^2)$ and $xF_3(x,Q^2)$ 
are given in leading (LO), next-to-leading (NLO) and next-to-next-to-leading order (NNLO) at different scales
of $Q^2$. The difference between LO and NLO turns out to be large since the NLO corrections are 
dominated by the 
gluon-$W$ fusion process, which contributes for the first time, and which reflects the  size of the 
gluon 
distribution.
They get much smaller comparing NLO and NNLO, where also the factorization scale uncertainty is 
expected to stabilize.
%----------------------------------------------------------------------------
\restylefloat{figure}
%----------------------------------------------------------------------------
\noindent
Here the ABM11 PDF set \cite{Alekhin:2012ig} at NNLO in the 3-flavor scheme was
used.  
%---------------------------------------------------------------------------------------------------------------
\section{Conclusions}
%---------------------------------------------------------------------------------------------------------------

\noindent
We reported on recent progress in the calculation of massive 3-loop operator matrix elements and 
Wilson Coefficients for deep-inelastic scattering for general values of the Mellin variable $N$.
Four years after a larger amount of Mellin moments for these quantities had been computed, six out of
eight OMEs and corresponding Wilson coefficients in the region $Q^2 \gg m^2$ have been calculated analytically. 
In parallel, quite a series of theoretical, mathematical and computer-algebraic technologies had to be 
newly developed and put significantly forward to make the present results possible. Here we would like to
note in particular the automated use of IBP-identities for massive 3-loop diagrams also containing 
local operator insertions in {\tt Reduze2} and modern summation technologies built in several advanced 
summation packages by the Linz-Group along with the development of the present project. These and 
related technologies are assumed to have a significant potential to be used in many other calculations 
in quantum field theory in the future. We also obtained a better insight into the calculation of 
the more difficult topologies, like those of $V$-graphs and the treatment of graphs with two massive 
fermion lines, being necessary to perform the forthcoming calculations. Furthermore, we obtained the 
asymptotic $O(\alpha_s^2)$ heavy-flavor corrections for deep-inelastic charged current scattering. 
The remaining part of the present project is still putting a series of very interesting challenges 
to be mastered.

%-------------------------------------------------------------------------------------------------

%------------------------------------------------------------------------------
%------------------------------------------------------------------------------
\end{document}